\documentclass[conference]{IEEEtran}

\usepackage{verbatim}
\usepackage{amsmath}
\usepackage{amssymb}
\usepackage{amsthm}
\usepackage{rotating}
\usepackage{algorithm}
\usepackage[noend]{algpseudocode}

\usepackage{tabularx}
\usepackage{subfigure}
\usepackage{graphicx}
\usepackage{url}
\usepackage{multirow}

\begin{document}
%
\title{Successful Nash Equilibrium Agent for a 3-Player Imperfect-Information Game}

\author{\IEEEauthorblockN{Sam Ganzfried}
\IEEEauthorblockA{Ganzfried Research\\
Miami Beach, Florida, 33139\\
Email: sam@ganzfriedresearch.com}
\and
\IEEEauthorblockN{Austin Nowak}
\IEEEauthorblockA{Florida International University\\
Miami, Florida, 33199\\
Email: anowak90@gmail.com}
\and
\IEEEauthorblockN{Joannier Pinales}
\IEEEauthorblockA{Florida International University\\
Miami, Florida, 33199\\
Email: jpina011@fiu.edu}
}

\maketitle

\begin{abstract}
Creating strong agents for games with more than two players is a major open problem in AI. Common approaches are based on approximating game-theoretic solution concepts such as Nash equilibrium, which have strong theoretical guarantees in two-player zero-sum games, but no guarantees in non-zero-sum games or in games with more than two players. We describe an agent that is able to defeat a variety of realistic opponents using an exact Nash equilibrium strategy in a 3-player imperfect-information game. This shows that, despite a lack of theoretical guarantees, agents based on Nash equilibrium strategies can be successful in multiplayer games after all.
\end{abstract}

\IEEEpeerreviewmaketitle

\section{Introduction}
\label{se:introduction}
Nash equilibrium has emerged as the central solution concept in game theory, in large part due to the pioneering PhD thesis of John Nash proving that one always exists in finite games~\cite{Nash50:Non-cooperative}. For two-player zero-sum games (i.e., competitive games where the winnings of one player equal the losses of the other player), the solution concept is particularly compelling, as it coincides with the concept of minimax/maximin strategies developed earlier by John von Neumann~\cite{Neumann28:Zur}. In that work von Neumann proved that playing such a strategy guarantees a value of the game for the player in the worst case (in expectation), and that the value is the best worst-case guarantee out of all strategies. Essentially this means that a player can guarantee winning (or at least tying) in the worst case if he follows such a strategy and alternates the role of player 1 and 2 in the game. So for two-player zero-sum games Nash equilibrium enjoys this ``unbeatability'' property. This has made it quite a compelling solution concept, and in fact agents based on approximating Nash equilibrium have been very successful, and have even been able to defeat the strongest humans in the world in the popular large-scale game of two player no-limit Texas hold 'em poker~\cite{Brown17b:Safe,Moravcik17b:DeepStack}. Nash equilibrium is additionally compelling for two-player zero-sum games due to the fact that it can be computed in polynomial time~\cite{Koller92:Complexity}.

For non-zero-sum games and games with more than two players, while Nash equilibrium is still guaranteed to exist due to Nash's result, none of these additional properties hold, as highlighted from the classic Battle of the Sexes game depicted in Figure~\ref{fi:battle}. This game has three Nash equilibrium strategy profiles: when both players select Opera (i.e., (Opera, Opera)), when both players select Football (Football, Football), and where both select their preferred option with probability $\frac{3}{5}$. Clearly in this case the success of playing a Nash equilibrium depends heavily on the strategy chosen by the other player. For example, if the wife follows her strategy from the first Nash equilibrium and plays Opera, but the husband follows his strategy from the second Nash equilibrium and plays Football, the wife will receive the worst possible payoff of 0 despite following a Nash equilibrium. While this example is just for a two-player game, the same phenomenon can occur in games with more than two players (though as described above it cannot occur in two-player zero-sum games). Even three-player zero-sum games are not special, as any two-player general-sum game can be converted into a three-player zero-sum game by adding a ``dummy'' third player whose payoff equals negative the sum of the other two players' payoff. Furthermore, even if we wanted to compute a Nash equilibrium
, it has been proven to be PPAD-complete and is widely conjectured that no efficient algorithm exists~\cite{Chen06:Settling,Chen05:Nash}, though several heuristic approaches have been developed for strategic-form games (i.e., matrix games such as Battle of the Sexes) with varying degrees of success in different settings~\cite{Berg17:Exclusion,Porter08:Simple,Govindan03:Global,Sandholm05:Mixed,Lemke64:Equilibrium}. There have also been techniques developed that approximate Nash equilibrium to a provably very small degree of approximation error in a 3-player imperfect-information game~\cite{Ganzfried08:Computing,Ganzfried09:Computing}.

\begin{figure}[!ht]
\begin{center}
\begin{tabular}{c|*{2}{c|}}
&Opera &Football \\ \hline
Opera &(3,2) &(0,0) \\ \hline
Football &(0,0) &(2,3) \\ \hline
\end{tabular}
\caption{Battle of the sexes game.}
\label{fi:battle}
\end{center}
\end{figure}

Thus, the problem of how to create strong agents for non-zero-sum and multiplayer games, and in particular the question of whether Nash equilibrium strategies are successful, remains an open problem---perhaps the most important one at the intersection of artificial intelligence and game theory. Of course, the most successful approach would not just simply follow a solution concept and would also attempt to learn and exploit weaknesses of the opponents~\cite{Johanson07:Computing,Ganzfried11:Game}. (Note that this would be potentially very helpful for two-player zero-sum games as well, as Nash equilibrium may not fully exploit mistakes of suboptimal opponents as much as successful exploitative agents even for that setting.) However, successfully performing opponent exploitation is very difficult, particularly in very large games where the number of game iterations and observations of the opponents' play is small compared to the number of game states. And furthermore, such approaches are susceptible to being deceived and counterexploited by sophisticated opponents. It is clear that pure exploitation approaches are insufficient to perform well against a mix of opponents of unknown skill level, and that a strong strategy rooted in game-theoretic foundations is required. 

The strongest existing agents for large multiplayer games have been based on approaches that attempt to approximate Nash equilibrium strategies~\cite{Abou10:Using,Gibson14:Regret}. In particular, they apply the counterfactual regret minimization algorithm~\cite{Zinkevich07:Regret,Lanctot09:Monte}, which has also been used for two-player zero-sum games and has resulted in super-human level play for both limit Texas hold 'em~\cite{Bowling15:Heads-up} and no-limit Texas hold 'em~\cite{Brown17b:Safe,Moravcik17b:DeepStack}. These agents have performed well in the 3-player limit Texas hold 'em division of the Annual Computer Poker Competition which is held annually at the AI conferences AAAI or IJCAI~\cite{ACPC}. Counterfactual regret minimization is an iterative self-play algorithm that is proven to converge to Nash equilibrium in the limit for two-player zero-sum games. It can be integrated with various forms of Monte Carlo sampling in order to improve performance both theoretically and in practice. For multiplayer and non-zero-sum games the algorithm can also be run, though the strategies computed are not guaranteed to form a Nash equilibrium. It was demonstrated that it does in fact converge to an $\epsilon$-Nash equilibrium (strategy profile in which no agent can gain more than $\epsilon$ by deviating) in the small game of 3-player Kuhn poker, while it does not converge to equilibrium in Leduc hold 'em~\cite{Abou10:Using}. It was subsequently proven that it guarantees converging to a strategy that is not dominated and does not put any weight on iteratively weakly-dominated actions~\cite{Gibson14:Regret}. While for some small games this guarantee can be very useful (e.g., for two-player Kuhn poker a high fraction of the actions are iteratively-weakly-dominated), in many large games (such as full Texas hold 'em) only a very small fraction of actions are dominated and the guarantee is not useful. Other approaches based on integrating the fictitious play algorithm with MDP-solving algorithms such as policy iteration have been demonstrated experimentally to converge to $\epsilon$-equilibrium for very small $\epsilon$ in a no-limit Texas hold 'em poker tournament endgame~\cite{Ganzfried08:Computing,Ganzfried09:Computing}. It has been proven that if these algorithms converge, then the resulting strategy profile constitutes a Nash equilibrium (while CFR does not have such a guarantee); however, the algorithms are not proven to converge in general, despite the fact that they did for the game that was experimented on.

The empirical success of the 3-player limit Texas hold 'em agents in the Annual Computer Poker Competition suggests that CFR-based approaches which are attempting to approximate Nash equilibrium are promising for multiplayer games. However, the takeaway is not very clear. First, the algorithms are not guaranteed to converge to equilibrium for this game, and there is no guarantee on whether the strategies used by the agents constitute a Nash equilibrium or are even remotely close to one. Furthermore, there were only a small number of opposing agents submitted to the competition who may have questionable skill level, so it is not clear whether the CFR-based approaches actually produce high-quality strategies or whether they just produced strategies that happened to outperform mediocre opponents and would have done very poorly against strong ones. While these CFR-based approaches are clearly the best so far and seem to be promising, they do not conclusively address the question of whether Nash equilibrium strategies can be successful in practice in interesting multiplayer games against realistic opponents.

In this paper we create an agent based on an exact Nash equilibrium strategy for the game of 3-player Kuhn poker. While this game is relatively small, and in particular quite small compared to 3-player limit Texas hold 'em, it is far from trivial to analyze, and has been used as a challenge problem at the Annual Computer Poker Competition for the past several years~\cite{ACPC}. A benefit of experimenting on a small problem is that exact Nash equilibrium strategies can be computed analytically~\cite{Szafron13:Parametrized}. That paper computed an infinite family of Nash equilibrium strategies, though it did not perform experiments to see how they performed in practice against realistic opponents. The poker competition also did not publish any details of the agents who participated, so it is unclear what approaches were used by the successful agents. We ran experiments with our equilibrium agent against 10 agents that were created recently as part of a class project. 
These agents were computed using a wide range of approaches, which included deep learning, opponent modeling, rule-based approaches, as well as game-theoretic approaches. We show that an approach based on using a natural Nash equilibrium strategy is able to outperform all of the agents from the class. This suggests that agents based on using Nash equilibrium strategies can in fact be successful in multiplayer games, despite the fact that they do not have a worst-case theoretical guarantee. Of course since we just experimented on one specific game there is no guarantee that this conclusion would apply beyond this to other games, and more extensive experiments would be needed to determine whether this conclusion would generalize.

\section{Three-player Kuhn poker}
\label{se:kuhn}
Three-player Kuhn poker is a simplified form of limit poker that has been used as a testbed game in the AAAI Annual Computer Poker Competition for several years~\cite{ACPC}. There is a single round of betting. Each player first antes a single chip and is dealt a card from a four-card deck that contains one Jack (J), one Queen (Q), one King (K), and one Ace (A). The first player has the option to \emph{bet} a fixed amount of one additional chip (by contrast in \emph{no-limit} games players can bet arbitrary amounts of chips) or to \emph{check} (remain in the hand but not bet an additional chip). When facing a bet, a player can \emph{call} (i.e., match the bet) or \emph{fold} (forfeit the hand). No additional bets or raises beyond the additional bet are allowed (while they are allowed in other common poker variants such as Texas hold 'em, both for the limit and no-limit variants). If all players but one have folded, then the player who has not folded wins the \emph{pot}, which consists of all chips in the middle. If more than one player have not folded by the end there is a \emph{showdown}, at which the players reveal their private card and the player with the highest card wins the entire pot (which consists of the initial antes plus all additional bets and calls). The ace is the highest card, followed by king, queen, and jack. As one example of a play of the game, suppose the players are dealt Queen, King, Ace respectively, and player 1 checks, then player 2 checks, then player 3 bets, then player 1 folds, then player 2 calls; then player 3 would win a pot of 5, for a profit of 3 from the amount he started the hand with. 

Note that despite the fact that 3-player Kuhn poker is only a synthetic simplified form of poker and is not actually played competitively, it is still far from trivial to analyze, and contains many of the interesting complexities of popular forms of poker such as Texas hold 'em. First, it is a game of imperfect information, as players are dealt a private card that the other agents do not have access to, which makes the game more complex than a game with perfect information that has the same number of nodes. Despite the size, it is not trivial to compute Nash equilibrium analytically, though recently an infinite family of Nash equilibria has been computed~\cite{Szafron13:Parametrized}. The equilibrium strategies exhibit the phenomena of \emph{bluffing} (i.e., sometimes betting with weak hands such as a Jack or Queen), and \emph{slow-playing} (aka \emph{trapping}) (i.e., sometimes checking with strong hands such as a King or Ace in order to induce a bet from a weaker hand). To see why, suppose an agent X played a simple strategy that only bet with an Ace or sometimes a King. Then the other agents would only call the bet if they had an Ace, since otherwise they would know they are beat (since there is only one King in the deck, if they held a King they would know that player X held an Ace). But now if the other agents are only calling with an Ace, it is unprofitable for player X to bet with a King, since he will lose an additional chip whenever another player holds an Ace, and will not get a call from a worse hand; it would be better to check and then potentially call with hopes that the other player is bluffing (or to fold if you think the player is bluffing too infrequently). A better strategy may be to bet with an Ace and to sometimes bet with a Jack as a bluff, to put the other players in a challenging situation when holding a Queen or King. However, player X may also want to sometimes check with an Ace as well so that he can still have some strong hands after he checks and the players are more wary of betting into him after a check.

A full infinite family of Nash equilibria for this game has been computed and can be seen in the tables from a recent article by Szafron et al.~\cite{Szafron13:Parametrized}. The family of equilibria is based on several parameter values, which once selected determine the probabilities for the other portions of the strategies. One can see from the table that randomization and including some probability on trapping and bluffing are essential in order to have a strong and unpredictable strategy. Thus, while this game may appear quite simple at first glance, analysis is still very far from simple, and the game exhibits many of the complexities of far larger games that are played competitively by humans for large amounts of money. 

\section{Nash equilibrium-based agent}
\label{se:agent}
One way wonder why it is worthwhile to create agents and experiment on three-player Kuhn poker, given that the game has been ``solved,'' as described in the preceding section. First, as described there are infinitely many Nash equilibria in this game (and furthermore there may be others beyond those in the family computed in the prior work). So even if we wanted to create an agent that employed a Nash equilibrium ``solution,'' it would not be clear which one to pick, and the performance would depend heavily on the strategies selected by the other agents (who may not even be playing a Nash equilibrium at all). This is similar to the phenomenon described for the Battle of the Sexes Game in the introduction, where even though the wife may be aware of all the equilibria, if she attends the Opera as part of the (O,O) equilibrium while the husband does football as part of the (F,F) equilibrium, both players obtain very low payoff despite both following equilibrium. A second reason is that, as also described in the introduction, Nash equilibrium has no theoretical benefits in three-player games, and it is possible that a non-equilibrium strategy (particularly one that integrates opponent modeling and exploitation) would perform better, even if we expected the opponents may be following a Nash equilibrium strategy, but particularly if we expect them to be playing predictably and/or making mistakes.

So despite that the fact that exact Nash equilibrium strategies have been computed for this game, it is still very unclear what a good approach is for creating a strong agent against a pool of unknown opponents. 

For our agent we have decided to use a Nash equilibrium strategy that has been singled out as being more robust than the others in prior work and that obtains the best worst-case payoff assuming that the other agents are following one of the strategies given by the computed infinite equilibrium family~\cite{Szafron13:Parametrized}. We depict this strategy in Table~\ref{ta:strategy1}. This table assigns values for the 21 free parameters in the infinite family of Nash equilibrium strategies. To define these parameters, $a_{jk}$, $b_{jk}$, and $c_{jk}$ denote the action probabilities for players $P_1$, $P_2$, and $P_3$ respectively when holding card $j$ and taking an aggressive action (Bet (B) or Call (C)) in  situation $k$, where the betting situations are defined in Table~\ref{ta:betting}.
Prior work has actually singled out a range of strategies that receive the best worst-case payoff; above we have described the lower bound of this space, and we also experiment using the strategy that falls at the upper bound (Table~\ref{ta:strategy2}).

\begin{table}[!ht]
\centering
\scriptsize
\begin{tabular}{|*{3}{c|}} \hline
P1 & P2 & P3\\ \hline
$a_{11} = 0$ & $b_{11} = 0$ & $c_{11} = 0$ \\ \hline
$a_{21} = 0$ & $b_{21} = 0$ & $c_{21} = \frac{1}{2}$ \\ \hline
$a_{22} = 0$ & $b_{22} = 0$ & $c_{22} = 0$ \\ \hline
$a_{23} = 0$ & $b_{23} = 0$ & $c_{23} = 0$ \\ \hline
$a_{31} = 0$ & $b_{31} = 0$ & $c_{31} = 0$ \\ \hline
$a_{32} = 0$ & $b_{32} = 0$ & $c_{32} = 0$ \\ \hline
$a_{33} = \frac{1}{2}$ & $b_{33} = \frac{1}{2}$ & $c_{33} = \frac{1}{2}$ \\ \hline
$a_{34} = 0$ & $b_{34} = 0$ & $c_{34} = 0$ \\ \hline
$a_{41} = 0$ & $b_{41} = 0$ & $c_{41} = 1$ \\ \hline
\end{tabular}
\caption{Parameter values used for our Nash equilibrium agent}
\label{ta:strategy1}
\end{table}

\begin{table}[!ht]
\centering
\scriptsize
\begin{tabular}{|*{3}{c|}} \hline
P1 & P2 & P3\\ \hline
$a_{11} = 0$ & $b_{11} = \frac{1}{4}$ & $c_{11} = 0$ \\ \hline
$a_{21} = 0$ & $b_{21} = \frac{1}{4}$ & $c_{21} = \frac{1}{2}$ \\ \hline
$a_{22} = 0$ & $b_{22} = 0$ & $c_{22} = 0$ \\ \hline
$a_{23} = 0$ & $b_{23} = 0$ & $c_{23} = 0$ \\ \hline
$a_{31} = 0$ & $b_{31} = 0$ & $c_{31} = 0$ \\ \hline
$a_{32} = 0$ & $b_{32} = 1$ & $c_{32} = 0$ \\ \hline
$a_{33} = \frac{1}{2}$ & $b_{33} = \frac{7}{8}$ & $c_{33} = 0$ \\ \hline
$a_{34} = 0$ & $b_{34} = 0$ & $c_{34} = 1$ \\ \hline
$a_{41} = 0$ & $b_{41} = 1$ & $c_{41} = 1$ \\ \hline
\end{tabular}
\caption{Parameter values used for our second Nash equilibrium agent}
\label{ta:strategy2}
\end{table}

\begin{table}[!ht]
\centering
\scriptsize
\begin{tabular}{|*{4}{c|}} \hline
Situation & P1 & P2 & P3\\ \hline
1 & -- & K & KK \\ \hline
2 & KKB & B & KB \\ \hline
3 & KBF & KKBF & BF \\ \hline
4 & KBC & KKBC & BC \\ \hline
\end{tabular}
\caption{Betting situations in three-player Kuhn poker}
\label{ta:betting}
\end{table}

\section{Experiments}
\label{se:experiments}
We experimented against 10 of 11 agents submitted recently for a class project (we ignored one agent that ran very slowly, which performed poorly). 
These agents utilized a wide variety of approaches, ranging from neural networks to counterfactual regret minimization to opponent modeling to rule-based approaches. For each grouping of 3 agents we ran matches consisting of 3000 hands between each of the 6 permutations of the agents (with the same cards being dealt for the respective positions of the agents in each of the duplicated matches). The number of hands per match (3000) is the same value used in the Annual Computer Poker Competition, and the process of duplicating the matches with the same cards between the different agent permutations is a common approach that significantly reduces the variance. We ran 10 matches for each permutation of 3 agents. Table~\ref{ta:comparison} shows the overall payoff (divided by 100,000) for each agent. The Nash agent received highest payoff. The results are very similar when using the upper and lower bound equilibrium strategies with the upper bound performing slightly better. 

\renewcommand{\tabcolsep}{3pt}
\begin{table}[!ht]
\centering
\scriptsize
\begin{tabular}{|*{11}{c|}} \hline
Nash &A1 &A2 &A3 &A4 &A5 &A6 &A7 &A8 &A9 &A10\\ \hline
2.81 (UB) &2.25 &1.18 &2.54 &-1.65 &2.32 &1.74 &-1.34 &-9.56 &-3.48 &1.42\\ \hline
2.81 (LB) &2.24 &1.17 &2.54 &-1.66 &2.32 &1.74 &-1.34 &-9.54 &-3.47 &1.42\\ \hline
\end{tabular}
\caption{Experiments using Nash agents against class project agents}
\label{ta:comparison}
\end{table}
\normalsize

\section{Conclusion}
\label{se:conclusion}
Creating strong agents for games with more than two players, and in particular the question of whether Nash equilibrium strategies are successful, is an important open problem---perhaps the most important one at the intersection of game theory and AI. We demonstrated that an agent based on following an exact Nash equilibrium is able to outperform agents submitted for a recent class project that utilize a wide variety of approaches. This suggests that agents based on using Nash equilibrium strategies can in fact be successful in multiplayer games, despite the fact that they do not have a worst-case theoretical guarantee. 

\bibliographystyle{IEEEtran}
\bibliography{D://FromBackup/Research/refs/dairefs}

\end{document}